\documentstyle[prl,aps,epsf,floats]{revtex}
\begin{document}
\newcommand{\beq}{\begin{equation}}
\newcommand{\eeq}{\end{equation}}
\newcommand{\beqn}{\begin{eqnarray}}
\newcommand{\eeqn}{\end{eqnarray}}
\newcommand{\bmath}{\begin{mathletters}}
\newcommand{\emath}{\end{mathletters}}
%\draft
\twocolumn[\hsize\textwidth\columnwidth\hsize\csname @twocolumnfalse\endcsname
\title{
Where is $99\%$ of the condensation energy of $Tl_2Ba_2CuO_y$ coming from?}
\author{J. E. Hirsch$^{a}$ and F. Marsiglio$^{b}$ }
\address{$^{a}$Department of Physics, University of California, San Diego,
La Jolla, CA 92093-0319\\
$^{b}$Department of Physics, University of Alberta, Edmonton,
Alberta, Canada T6G2J1}
 
\date{\today} 
\maketitle 
\begin{abstract} 
Anderson's interlayer tunneling model can account for up to $1\%$ of the
condensation energy of $Tl_2Ba_2CuO_y$. Here we account for
the remaining $99\%$. We predict an in-plane kinetic energy gain of 1 to 3 meV 
per planar oxygen
in the optimally doped material when it goes superconducting. It is suggested that 
the effect may be most easily detected in underdoped dirty samples.
\end{abstract}
\vskip2pc]

In conventional superconductors, the opening of the superconducting
energy gap causes a reduction in optical absorption at low
frequencies (below twice the gap value)\cite{gt}, and this missing
spectral weight is transfered to the zero-frequency $\delta-$function that
determines the penetration depth\cite{fgt} (so-called 
Ferrell-Glover-Tinkham (FGT) sum rule). It was proposed in Ref. 3 that in certain
superconductors where pairing would arise from lowering of kinetic rather
than of potential energy, the FGT sum rule would in appearance be violated,
leading to the following novel phenomena: (1) a London penetration
depth shorter than would be expected from the low frequency change
in optical absorption and/or the normal state effective mass\cite{lond},
 and (2) a decrease in optical absorption
at frequencies much higher than, and unrelated to, the superconducting energy
gap, possibly near infrared or visible. Recent experimental observations 
by Basov and coworkers in several high $T_c$ oxides\cite{bas1} provide striking 
evidence for the
existence of this physics in high $T_c$ cuprates: for c-axis light
polarization, the change in low frequency optical spectral weight
accounts for just about $50\%$ of the total spectral
weight in the zero-frequency $\delta-$function, indicating that the remaining
spectral weight comes from frequencies higher than the highest
frequencies reached in the experiment, namely $150$ meV. Note that this
energy is about 6 times the value of the gap, within which all the optical 
spectral weight change should be contained according to conventional
BCS theory.\cite{bas1} These findings are also consistent with earlier
experiments by Fugol et al\cite{fugo} that report an anomalous decrease in optical
absorption at frequencies in the visible range when high $T_c$ samples
become superconducting.

The well-known interlayer tunneling theory\cite{ilt} (ILT) of the cuprates
predicts a lowering of the c-axis kinetic energy as the system goes
superconducting, so that the considerations of Ref. 3 apply to that
model for the case when light is polarized along the c direction.
Realizing this, Chakravarty\cite{chak} applied the analysis of Ref. 3
to the particular case of c-axis conductivity and found (not surprisingly)
that a lowering of kinetic energy in that direction would be consistent with an 
apparent violation of the FGT sum rule for c-axis light polarization, and hence 
the phenomena discussed above. For planar light polarization instead,
no apparent sum rule violation is expected within the ILT model. 

However, within the ILT theory the kinetic energy lowering should entirely account 
for the condensation
energy of the superconductor\cite{and1,legg,tsve}. In fact, in the analysis of
c-axis transport of various high $T_c$ oxides where agreement with ILT theory is
claimed\cite{and1}, e.g. $La_{2-x}Sr_xCuO_4$, no distinction is made as to which 
frequency range
in the optical absorption the $\delta$-function spectral weight is coming from:
the entire weight is attributed to kinetic energy lowering, which in turn
implies that ILT theory would also be consistent with observation of $no$ sum
rule violation \cite{pred}. 
However, the condensation energy of
optimally doped $Tl_2Ba_2CuO_y$ (Tl2201) is estimated from specific heat measurements
to be $80-100 \mu eV$ per $CuO_2$ unit\cite{lora}. 
Even in the extreme case 
where the entire weight in the superfluid condensate would come from
kinetic energy lowering the c-axis
penetration depth required would be less than $2\mu m$\cite{and1,legg}, 
which is significantly
smaller than the measured penetration depth in that material, 
$\sim 17 \mu m$\cite{tsve,schu,mole}. In other words, the measured penetration depth 
implies
that the kinetic energy of c-axis motion in the superconducting state of
$Tl_2Ba_2CuO_y$ accounts for less than $1\%$ of its condensation 
energy\cite{and1,legg,tsve,schu,mole}.

Still, we argue that Basov's and Fugol's experiments have clearly demonstrated that 
in high $T_c$ materials
at least some of the superconducting condensation energy originates in kinetic energy. 
It is then natural to ask whether
the rest might too. If so, the remaining condensation energy should arise from 
lowering of the $in-plane$ kinetic energy. However, Basov and coworkers report\cite{bas1}
no apparent sum rule violation in the in-plane response within experimental
accuracy so far, approximately $10\%$\cite{bas2}.

This is however not surprising. We argue that the reason a sum rule violation
has been detected in c-axis transport and not in in-plane transport so far is
simply because the materials are in the dirty limit for c-axis conduction and close 
to the clean limit for in-plane conduction\cite{mare}. Note that the kinetic energy 
associated with the measured in-plane penetration depth of Tl2201 of
$\sim 2000A$ is of order $20-60 meV$ per $CuO_2$ unit, of which the
condensation energy is only a tiny fraction. Still, as we discuss below, we
expect that an in-plane sum rule violation will be detectable.

Within the model of hole superconductivity\cite{hole}, pairing arises from 
lowering of kinetic energy $in$ $all$ $directions$ when the system goes superconducting.
Thus, the model predicts optical sum rule violation both in the in-plane and
in the inter-plane response\cite{appa}. Because in-plane couplings are much larger than 
interplane ones, the bulk of the condensation energy in the model arises from
kinetic energy lowering in the plane. Nevertheless, as we show in this paper, the model
can readily account for the observed large sum rule violation in the
c-direction and a much smaller sum rule violation in the plane, consistent
with existing experimental results. It is predicted that with increased experimental
accuracy an in-plane sum rule violation will be detected, most easily in
the underdoped regime, which should readily account for the remaining
$99\%$ of the condensation energy in $Tl_2Ba_2CuO_y$ and other high $T_c$ cuprates. 
In fact, the model predicts a lowering of kinetic energy that is much larger than the 
condensation energy, that is partially offset by an increase in potential
energy.

Anderson remarks\cite{and1} that for several high $T_c$ materials other than
$Tl_2Ba_2CuO_y$ the condensation
energy is completely accounted for by the lowering of the c-axis kinetic
energy as determined from the c-axis penetration depth, and concludes that
for those cases ``this agreement effectively rules out any intralayer theory
of high $T_c$''. This conclusion is logically flawed: the observations are
entirely compatible with the existence of a much larger kinetic energy lowering
from in-plane motion together with an increase in potential energy upon pairing,
as predicted by our model.

The frequency-dependent conductivity in the superconducting state is
given by $\sigma_{1s}(\omega)=D\delta(\omega)+\sigma_{1s}^{reg}(\omega)$,
with $\sigma_{1s}^{reg}(\omega)$ the regular part. 
As discussed in Ref. 3, the superfluid weight in the zero-frequency 
$\delta-$function, $D$, is given by
\beq
D_\mu=\delta A_l^\mu+\delta A_h^\mu
\eeq
with $\delta A_l^\mu$ the low frequency missing area, due to intra-band 
transitions, and
\beq
\delta A_h^\mu=\frac{\pi e^2 a_\mu^2}{2\hbar ^2 v}[<-T_\mu>_s-<-T_\mu>_n]
\eeq
the missing area from much higher frequencies, related to interband
transitions. $v$ gives the volume of the unit cell,
$\mu$ indicates the direction of light polarization,
and the right side of Eq. (2) gives the change in the carrier's
kinetic energy in the $\mu$ direction as the system goes superconducting.
The London penetration depth is given by $\lambda_\mu=c/(8D_\mu)^{1/2}$, and
an apparent sum rule violation exists if it is shorter than what is
obtained from just the low frequency missing area, $\delta A_l^\mu$\cite{appa,lond},
indicating the existence of $\delta A_h^\mu$ even if it is not detected directly.
The degree of sum rule 'violation' can be characterized by the parameter
\beq
V_\mu=\frac{\delta A_h^\mu}{\delta A_l^\mu+\delta A_h^\mu}
\eeq
and was found by Basov et al to be approximately $0.5$ in the $c$ direction
for several high $T_c$ materials including $Tl2201$.

For application to ILT theory the normal state kinetic energy in the c direction
in Eq. (2) is assumed to be negligible, and Eq. (2) yields
\beq
\frac{1}{\lambda_c^2}=\frac{4\pi e^2 d}{\hbar^2c^2a^2}<-T_c>_s
\eeq
with $a$ and $d$ in-plane and interplane lattice constants. The negative of the
superconducting kinetic energy, $<-T_c>_s$, is equated to the condensation
energy, $\epsilon_{cond}$. For the case of $Tl_2Ba_2CuO_y$ $\lambda_c\sim 17\mu m$,
$d=11.6A$, $a=3.9A$, eq. (4) yields $\epsilon_{cond}=0.98 \mu eV$ per $CuO_2$ 
unit, two orders
of magnitude smaller than the value estimated from specific heat
measurements\cite{lora}.

In the model of hole superconductivity, the kinetic energy in the $\mu$ direction
is given by\cite{hole}
\beqn
T_\mu&=&\sum_i [t_\mu+\Delta t_\mu(n_{i,-\sigma}+n_{i+\mu,-\sigma})]
[c_{i\sigma}^\dagger c_{i+\mu,\sigma} +h.c.]\equiv \nonumber \\
& &T_\mu^t
+T_\mu^{\Delta t}
\eeqn
with $c_{i\sigma}^\dagger$ a hole creation operator. The effective hopping
in the $\mu$ direction is given by
\beq
t_\mu^{eff}=t_\mu+\Delta t_\mu n
\eeq
and it increases linearly with hole concentration $n$. To a very good
approximation\cite{appa} the change in kinetic energy Eq. (2) is simply
given by the expectation value of the correlated hopping term
\beq
\delta A_h^\mu=\frac{\pi e^2 a_\mu^2}{2\hbar ^2 v}<-T_\mu^{\Delta t}>_{s,a}
\eeq
where the subindex $a$ indicates that only the $anomalous$ expectation values 
are to be included\cite{appa}.

In the clean limit at zero temperature we have simply
\beq
\delta A_l^\mu=\frac{\pi e^2 a_\mu^2}{2\hbar ^2 v}<-T_\mu^t>_s
\eeq
where the average of the single particle kinetic energy includes also
the 'normal' expectation values of the $\Delta t$ term, yielding the
renormalized single particle hopping Eq. (6). The degree of sum rule
violation in that case is given by
\beq
V_\mu=\frac{<T_\mu^{\Delta t}>_{s,a}}
{<T_\mu^t>_s+<T_\mu^{\Delta t}>_{s,a}}.
\eeq

The full Hamiltonian for the model includes in addition to the
kinetic energy an on-site and nearest neighbor Coulomb repulsion:
\beq
H_{Coul}=H_U+H_V=U\sum_i n_{i\uparrow}n_{i\downarrow}
+V\sum_{<i,j>}n_in_j
\eeq
and the condensation energy of the superconductor is given by
\beqn
&\epsilon_{cond}&=[<T^t>_n-<T^t>_s]-
<T^{\Delta t}>_{s,a} - \nonumber \\
& &<H_U>_{s,a}-
<H_V>_{s,a}\equiv \Delta \epsilon _t+\epsilon_{\Delta  t}+
\epsilon_U+\epsilon_V
\eeqn
where the expectation values with subindex $a$ indicate that only the
anomalous contributions are to be included. In Eq. (11), a sum over the 
different directions is implicit for the terms that involve nearest
neighbors. Quite generally, all contributions to $\epsilon_{cond}$ other than
$\epsilon_{\Delta t}$ are negative: in the paired state, particles
are closer together and experience larger Coulomb repulsion, and the
single particle kinetic energy $<T_t>$ is optimal (most negative) in the 
normal state. A positive $\epsilon_{cond}$ originates from a large
negative contribution from $<T^{\Delta t}>_{s,a}$, i.e. a large kinetic
energy lowering due to correlated hopping, which is to some extent compensated 
by the other terms
in Eq. (11). In weak coupling, Eq. (11) is given by the usual form
\beq
\epsilon_{cond}=\frac{\Delta_0^2}{2}N(\epsilon_F)
\eeq
with $\Delta_0$ the energy gap and $N(\epsilon_F)$ the density of states 
at the Fermi energy.    

\begin{figure}
\centerline{
\epsfysize 3.0in
\epsfbox{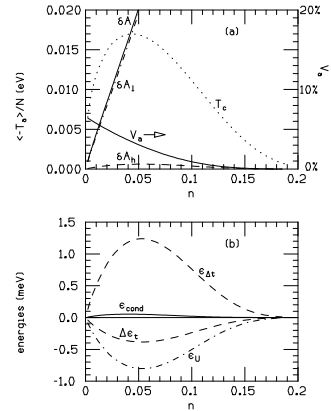}}
\vskip0.01cm
\caption{
 Average in-plane kinetic energy per lattice site at zero temperature
vs. hole concentration for case 1 parameters (given in the text). The solid, 
dot-dashed and
dashed lines give the total kinetic energy, single particle and pair
contributions respectively. The sum rule violation parameter $V_a$ in the clean
limit (Eq. (9)) is also shown as a solid line (right scale). The critical temperature
versus hole concentration, with maximum $T=85K$, is indicated by a dotted
line. (b) The various contributions to the condensation energy per O site (Eq. (11)) 
versus hole
concentration. 
}
\label{fig:fig1}
\end{figure}

In the model Eq. (5), each site denotes an oxygen atom in a Cu-O plane.
Consider the following parameters to describe the CuO planes of Tl2201 (case 1): 
$U=5eV$, $t=0.19 eV$,
$\Delta t=0.29 eV$. In the optimally doped case, with hole concentration
$n=0.045$ (per O atom), we obtain $T_c=85K$. Explicit calculation with the
expressions given in Ref. 18 yields the results shown in 
Fig.~\ref{fig:fig1}. The sum
rule violation parameter Eq. (9) is $V_a=3.4\%$ for the optimally doped
case, well below the current experimental accuracy. It increases up to 
$V_a=6.5\%$ in the underdoped regime. As we will show below however, the 
resulting sum rule violation in the $c$-direction can be much larger,
consistent with observations. The condensation energy per O
atom is $53 \mu eV$ in the optimally doped case, consistent with
observations ($\epsilon_{cond}\sim 100 \mu eV$ per $CuO_2$ unit). Note 
that it results from a much larger kinetic energy lowering due to 
correlated hopping, $\epsilon_{\Delta t}=1.2 meV$, which is partially
compensated by an increase in potential energy and in kinetic energy from
single particle hopping.

The penetration depth in the ab plane in the clean limit is given by
\beq
\frac{1}{\lambda_a^2}=\frac{4\pi e^2 }{\hbar^2c^2 d}
[<-T_a^t>_s+<-T_a^{\Delta t}>_{s,a}].
\eeq
For the optimally doped case, we find $<-T_a^t>_s=-17.60 meV$ and
$<-T_a^{\Delta t}>=-0.610meV$, and with  $d=11.6A$
 Eq. (13) yields
$\lambda_a=3697A$. However because there are other atoms between the
CuO layers which are involved in the superconductivity it may be more
appropiate to use $d\sim 3A$ in Eq. (13), which then yields
$\lambda_a=1880A$, consistent with experimental observations\cite{lama}.
Also, with the parameters chosen above, the density of states per $CuO_2$
unit is $N(\epsilon_F)=1.25 eV^{-1}$, consistent with experimental estimates
for Tl2201.

\begin{figure}
\centerline{
\epsfysize 3.0in
\epsfbox{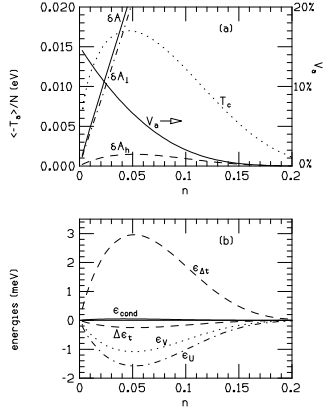}}
\vskip0.01cm
\caption{
Same as Figure 1 for case 2 parameters, with non-zero nearest neighbor
repulsion, given in the text. Note that the violation parameter $V_a$ as well
as the kinetic energy lowering $\epsilon_{\Delta t}$ are substantially
larger than in Fig. 1.
}
\label{fig:fig2}
\end{figure}

Similarly, Fig.~\ref{fig:fig2}  shows results when a nearest neighbor repulsion is
included (case 2). Here, $U=5eV$, $V=0.65 eV$, $t=0.19 eV$, and $\Delta t=
0.51 eV$ is chosen to again yield $T_c=85K$ for the optimally doped
case. Here the kinetic energy lowering is larger, and yields a
sum rule violation $V_a=7.3\%$ for the optimally doped
case which increases to 
$V_a=15\%$ in the underdoped regime. We expect the actual parameters 
describing Tl2201 to be somewhere in the range spanned by these two examples.
If the condensation energy was in fact somewhat larger than the current
estimates the model would require a smaller value of the parameter
$t$ which would lead to an even larger effect.

What about the c direction? The simplest assumption\cite{hole} is that $t$
and $\Delta t$ are reduced by the same factor. If we take $\lambda_c=17\mu m$ from
experiment\cite{mole}, and assume in Eq. (4) that the c-axis kinetic energy lowering
due to $\Delta t$ supplies $50 \%$ of the $\delta-$function weight, as found by
Basov (i.e. $50\%$ sum rule violation), 
we obtain $<T_c^{\Delta t}>=0.25\mu eV$ (per O atom). For the parameters
of case 1 above, we get an anisotropy factor
\beq
\frac{<T_a^{\Delta t}>}{<T_c^{\Delta t}>}=2440
\eeq
and the kinetic energy from single particle hopping scales approximately with the same
factor\cite{lond}. This corresponds to an anisotropy in hopping amplitudes
(or effective masses) $t_a/t_c=m^*_c/m^*_a \sim 50$\cite{lond}.
The penetration depth in the clean limit in the c direction is
given by
\beq
\frac{1}{(\lambda_c^{clean})^2}=\frac{4\pi e^2 d }{\hbar^2c^2 a^2}
[<-T_c^t>_s+<-T_c^{\Delta t}>_{s,a}]
\eeq
and yields $\lambda_c^{clean}=4.3 \mu m$. The observed penetration depth
in the c direction is however a factor of 4 larger, which we attribute to
the effect of disorder reducing the contribution of 
$\delta A_l$ in Eq.(1)\cite{marsiglio96}:
\beq
\delta A_l\sim \delta A_l^{clean}\times \pi\Delta_0/(\hbar/\tau)
\eeq
with an impurity
scattering rate $\hbar/\tau_c\sim 50\Delta _0$, well in the dirty limit.
In other words, the fact that the penetration depth we obtain from
Eq. (15) is much smaller than the observed one, assuming this is due to 
the effect of disorder, is consistent with the
observation that the sum rule violation in the c-direction is much
larger than in the plane. Quite generally,
under the assumption that low and high frequency missing areas have the
same ratio in all directions in the clean limit Eqs. (3) and (16) yield
\beq
\frac{\hbar/\tau_a}{\hbar/\tau_c}=\frac{\frac{1}{V_c}-1}{\frac{1}{V_a}-1}
\eeq
hence a larger scattering rate in the $c$ direction will necessarily
be associated with a larger sum rule violation in that direction.

As a further consistency check on the validity of this analysis we consider 
the normal states conductivities.
For $\Delta _0=13meV$ as obtained from our model, the above analysis
implies $\hbar/\tau_c=5240 cm^{-1}$.
For in-plane transport instead, Puchkov et al find\cite{puch} a scattering
rate $1/\tau_a=560 cm^{-1}$, a factor of $10$ smaller (which is consistent
with a much smaller in-plane sum rule violation), and 
$\sigma_n(\omega \rightarrow 0)\sim 5000 (\Omega cm)^{-1}$. Hence we expect
in the $c$ direction a conductivity that is smaller by a factor of approximately
$\tau _a/\tau _c \times m^*_c/m^*_a \sim 500 $,
i.e. $\sigma_n^c\sim 10 (\Omega cm)^{-1}$
 and frequency-independent
over several times $\Delta_0$. This is consistent with observations.

 Note that the c-axis kinetic energy lowering
for these parameters is only $0.5\%$ of the total condensation energy,
and only $0.02\%$ of the kinetic energy lowering originating in in-plane
motion. Results for the other set of parameters discussed above are
similar.

Will it be possible to observe a sum rule violation in the in-plane
response? As shown above, the sum rule violation increases in the 
underdoped regime and hence it may be possible that the in-plane violation
will be detected even if the system remains in the clean limit, especially
if the nearest neighbor repulsion is appreciable (case 2 above). Furthermore,
by introducing disorder, for example by ion irradiation\cite{bas3},
 it may be possible to reduce the contribution
of $\delta A_l$ to the superfluid density, moving the system towards the
dirty limit, thus increasing the violation
parameter.  In this connection it should also be noted that experimental 
results in the c direction show a faster variation of the violation 
parameter with doping than found in Figs. 1 and 2\cite{bas2}. This however is
simply explained by the fact that the system is moving towards the
clean limit with increased doping, as evidenced by the rapid increase in 
the optical conductivity, thus increasing the contribution of
$\delta A_l$ faster than given by the results in Figs. 1 and 2. Quantitative fits 
will be discussed elsewhere.

In summary, the model of hole superconductivity predicted an apparent
optical sum rule violation\cite{appa} long before it was experimentally observed,
which would be a manifestation of the reduction in effective mass and
consequent lowering of kinetic energy that occurs upon pairing. By
assuming that Tl2201 is in the clean limit for in-plane conduction and in
the dirty limit for the interplane response (which is consistent with
observations\cite{mare}), we found that the model can
explain experimental observations. In the optimally doped case, the
kinetic energy lowering from c-axis motion accounts for $\sim 1\%$ of the
condensation energy, and that of in-plane motion is 20-50 times larger
than the condensation energy. In particular, the model can account
for the $99\%$ condensation energy that is missing in the alternative
proposed explanation of the sum rule violation experiments, the ILT model\cite{ilt}.
The fact that the predicted in-plane kinetic energy lowering is much larger
than the condensation energy is fortunate and should allow for its
experimental detection.

\acknowledgements
The authors are grateful to D. Basov for stimulating discussion and for
sharing his experimental results prior to publication. Hospitality of the 
Aspen Center for Physics where this work was started is gratefully
acknowledged.


\begin{references}
\bibitem{pred} That is, the entire spectral weight in the
$\delta$-function could come from low frequencies
and the result would still be consistent with ILT theory as long as the weight in the
$\delta$-function matched the condensation energy.
Note that this implies in particular that ILT theory $could$ $not$
$have$ $predicted$ Basov's observation of apparent sum rule violation.
\end{references}
\end{document}